\documentclass[twocolumn,prl,amsmath,amssymb,floatfix,reprint,square,comma,numbers]{revtex4-1}

\usepackage[latin9]{inputenc}
\usepackage{textcomp}
\usepackage{amstext}
\usepackage[unicode=true,
 bookmarks=false,
 breaklinks=false,pdfborder={0 0 1},colorlinks=false]
{hyperref}

\usepackage{graphicx}
\usepackage{dcolumn}
\usepackage{bm}
\usepackage{multirow}
\usepackage{enumitem}
\usepackage{color}
\usepackage{braket}

\begin{document}
\title{Generation and controllable switching of superradiant and subradiant states in a 10-qubit superconducting circuit}

\author{Zhen Wang$^{1}$}
\author{Hekang Li$^{2}$}
\author{Wei Feng$^{1}$}
\author{Chao Song$^{1}$}
\author{Wuxin Liu$^{1}$}
\author{Qiujiang Guo$^{1}$}
\author{Xu Zhang$^{1}$}
\author{Hang Dong$^{1}$}
\author{Dongning Zheng$^{2,3}$}
\email{dzheng@iphy.ac.cn}
\author{H. Wang$^{1}$}
\email{hhwang@zju.edu.cn}
\author{Da-Wei Wang$^{1,3}$}
\email{dwwang@zju.edu.cn}
\affiliation{$^{1}$ Interdisciplinary Center for Quantum Information and Zhejiang Province Key Laboratory \\ of Quantum Technology and Device,
Department of Physics and State Key Laboratory \\ of Modern Optical Instrumentation, Zhejiang University, Hangzhou 310027, China,\\
$^2$ \mbox{Institute of Physics, Chinese Academy of Sciences, Beijing 100190, China},
$^3$ CAS Center for Excellence in Topological Quantum Computation, University of Chinese Academy of Sciences, Beijing 100190, China
}

\date{\today}
\begin{abstract}
Superradiance and subradiance concerning enhanced and inhibited collective radiation of an ensemble of atoms have been a central topic in quantum optics. However, precise generation and control of these states remain challenging. Here we deterministically generate up to 10-qubit superradiant and 8-qubit subradiant states, each containing a single excitation, in a superconducting quantum circuit with multiple qubits interconnected by a cavity resonator. The $\sqrt{N}$-scaling enhancement of the coupling strength between the superradiant states and the cavity is validated. By applying appropriate phase gate on each qubit, we are able to
switch the single collective excitation between superradiant and subradiant states. While the subradiant states containing a single excitation are forbidden from emitting photons, we demonstrate that they can still absorb photons from the resonator. However, for even number of qubits, a singlet state with half of the qubits being excited can neither emit nor absorb photons, which is verified with 4 qubits. This study is a step forward in coherent control of collective radiation and has promising applications in quantum information processing.
\end{abstract}

\maketitle

Since Dicke's seminal paper in 1954 \cite{1954_PR_DickePropose},
superradiance featuring the enhanced cooperative radiation of atoms
has always been a research focus in quantum optics. Apart from intriguing
physics such as the superradiant phase transitions \cite{K. Hepp,key-3},
superradiance has promising applications in quantum communication
\cite{key-4,key-5}, ultra-narrow-linewidth superradiant lasers \cite{key-6,key-7},
and sensitive gravimeters \cite{key-8}. The cooperativity between
the atoms can also lead to subradiance \cite{key-9}, the inhibited
collective radiation closely related to the decoherence-free subspaces
\cite{key-10,key-11}. By taking advantage of their radiation characteristics,
superradiance is perfect for fast writing and reading of quantum information
while subradiance can be used in quantum memory \cite{key-12}. This
application in quantum information processing requires a fast switching
between superradiance and subradiance.

Although superradiance has been demonstrated in many physical systems
including hot \cite{key-13,key-14} and cold atoms \cite{key-15,key-16,key-17,key-18,key-19,key-20},
trapped ions \cite{key-21,key-22}, nuclear X-ray radiation \cite{key-23},
and superconducting circuits \cite{key-24,key-25,key-26}, precise
control of superradiant states and switching from superradiant to
subradiant states remain challenging. For artificial atoms such as
superconducting qubits with advantagous controlability, the maximum
number of qubits demonstrated in superradiance is merely four \cite{key-25,key-27}.
Compared with superradiance, the experimental realization of subradiance
is even more difficult due to their weak coupling with photons. In
atomic ensembles only partial subradiance with a weak signal has been
observed \cite{key-28}. In trapped ions and superconducting circuits
the maximum number of atoms in subradiant states is two \cite{key-21,key-22,key-29}.
Despite that the subradiant subspaces were already shown in Dicke's
original paper, the quantum transition in subradiant subspaces and
the radiation properties of subradiant states containing more than
one excitations have never been experimentally tested.

\begin{figure}
\begin{center}
\includegraphics[width=3.4in]{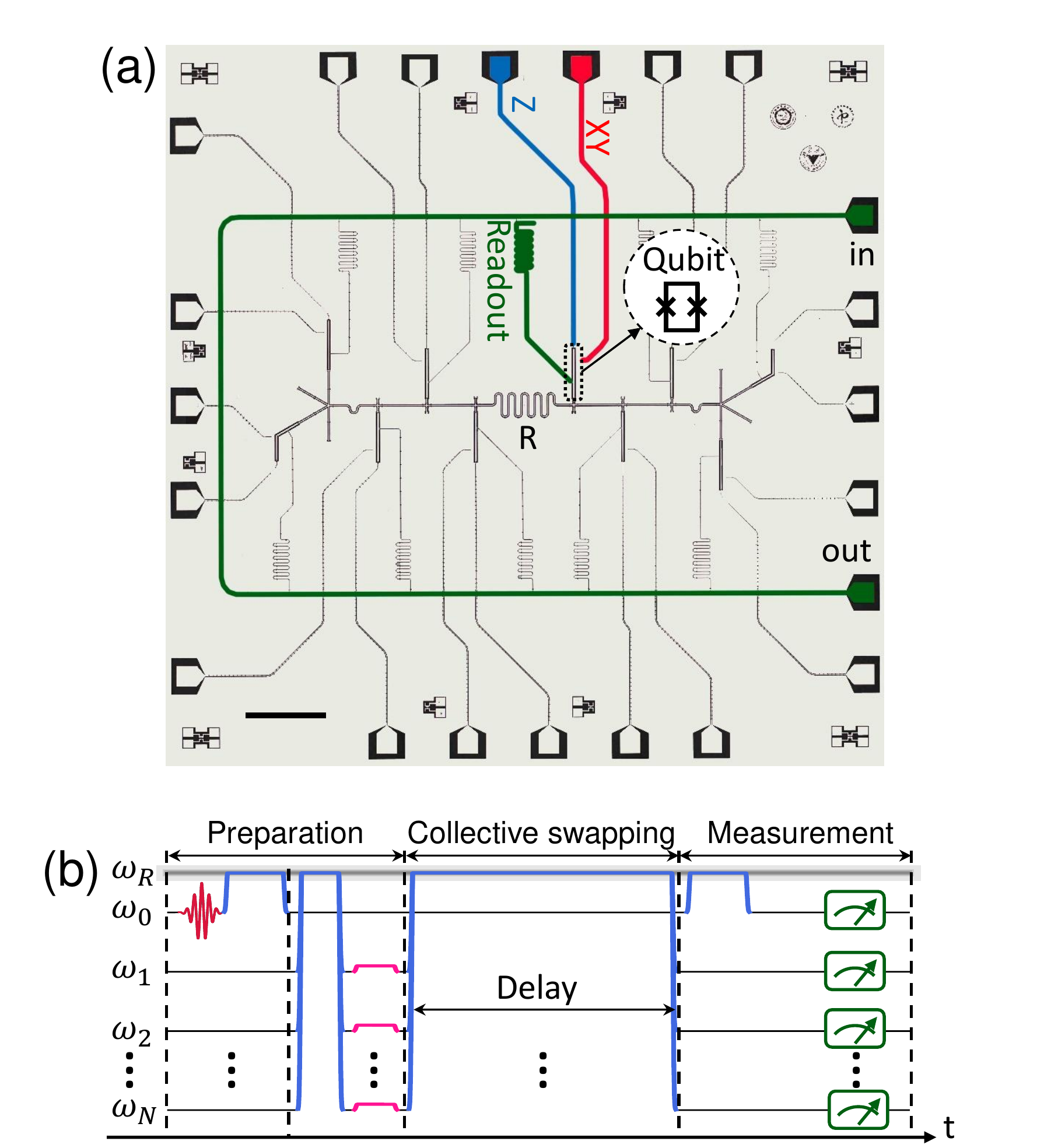}
\caption{\label{figure1}
(a) Image of the device where 10 qubits, shown as line shapes, capacitively couple to the central bus resonator. Each qubit has a Z line for frequency biasing, an XY line for microwave excitation, and a readout resonator for qubit state measurement. A ten-tone microwave signal, which targets the resonance frequencies of all readout resonators, is passed through the launch pads ``in'' and ``out'' as labeled for the multiqubit readout. The scale bar is 1 mm.
(b) Pulse sequences used to generate and characterize the single-excitation superradiant/subradiant states, where horizontal axis is for time and verticle axis represents frequency. For Q$_0$ with frequency $\omega_0$, the sinusoid with a Gaussian envelope is a microwave pulse acting as a $\pi$-rotational gate. 
The two rectangular Z pulses with durations of $t_\textrm{iSWAP}$ ($\approx \pi/2g$) are iSWAP gates: The first one transfers an excitation quantum from Q$_0$ to $R$
and the second one is for measurement of the resonator population.
For Q$_1$ to Q$_N$, the big rectangular Z pulses (blue) simultaneously bring all qubits into resonance with $R$: The first set has a duration of $\approx\pi/2\sqrt{N}g$ 
while the second set takes duration values continuously from 0 to 100 ns for observation of the collective swapping dynamics between the $N$ qubits and $R$. The small rectangular Z pulses (red) are phase gates with durations of 100~ns, whose combination determines whether the prepared state is in superradiance or subradiance.}
\end{center}
\end{figure}

In this Letter, we demonstrate the generation of superradiant states
with up to 10 qubits and subradiant states with up to 8 qubits in
a superconducting quantum circuit, where multiple qubits are directly
coupled to a common bus resonator ($R$). The superradiance is generated
by a cooperative absorption of a photon from resonator $R$ and is
characterized by the simultaneous readout of all qubits and the resonator,
which yields Rabi oscillations validating the factor $\sqrt{N}$-enhanced
coupling strength between the collective qubits and the resonator.
By applying appropriate single-qubit phase gates, we demonstrate a
controllable switching between superradiance and subradiance, and
show that the collective coupling can be enhanced or inhibited at
will. While the single-excitation subradiant states cannot radiate
photons into the resonator, we experimentally show that they can absorb
another photon from the resonator. It is well known that the singlet
states of even number of spin excitations can be used in noiseless
quantum codes \cite{key-10,key-11}. We realize such a singlet subradiant
state containing 2 excitations of 4 qubits and verify that it can
neither emit nor absorb photons from the resonator.




The qubit-resonator interaction Hamiltonian is described by the Tavis-Cummings
model \cite{key-30} under the rotating-wave approximation,
\begin{equation}
\frac{H}{\hbar}=\omega_{R}{a}^{\dag}{a}+\sum_{j=0}^{9}\omega_{j}\sigma_{j}^{+}\sigma_{j}^{-}+\sum_{j=0}^{9}g_{j}(\sigma_{j}^{+}{a}+\sigma_{j}^{-}{a}^{\dag}),\label{eqn:hamiltionan}
\end{equation}
where ${a}^{\dag}$ (${a}$) is the creation (annihilation) operator
of resonator $R$ with a fixed frequency of $\omega_{R}/2\pi\approx$
5.69 GHz, $\sigma_{j}^{+}$ ($\sigma_{j}^{-}$) is the raising (lowering)
operator of qubit $Q_{j}$ with a frequency $\omega_{j}/2\pi$ tunable
in the range from 5 to 6 GHz, and $g_{j}$ is the $Q_{j}$-$R$ coupling
strength, which can be approximated as $g\equiv\sum_{j=0}^{9}{g_{j}}/10\approx13.5$~MHz
since all qubit-resonator coupling strength values are measured to
be close. See Fig.~1(a) and Supplementary Material for more details.

\begin{figure}[t]
\begin{center}
\includegraphics[width=3.4in]{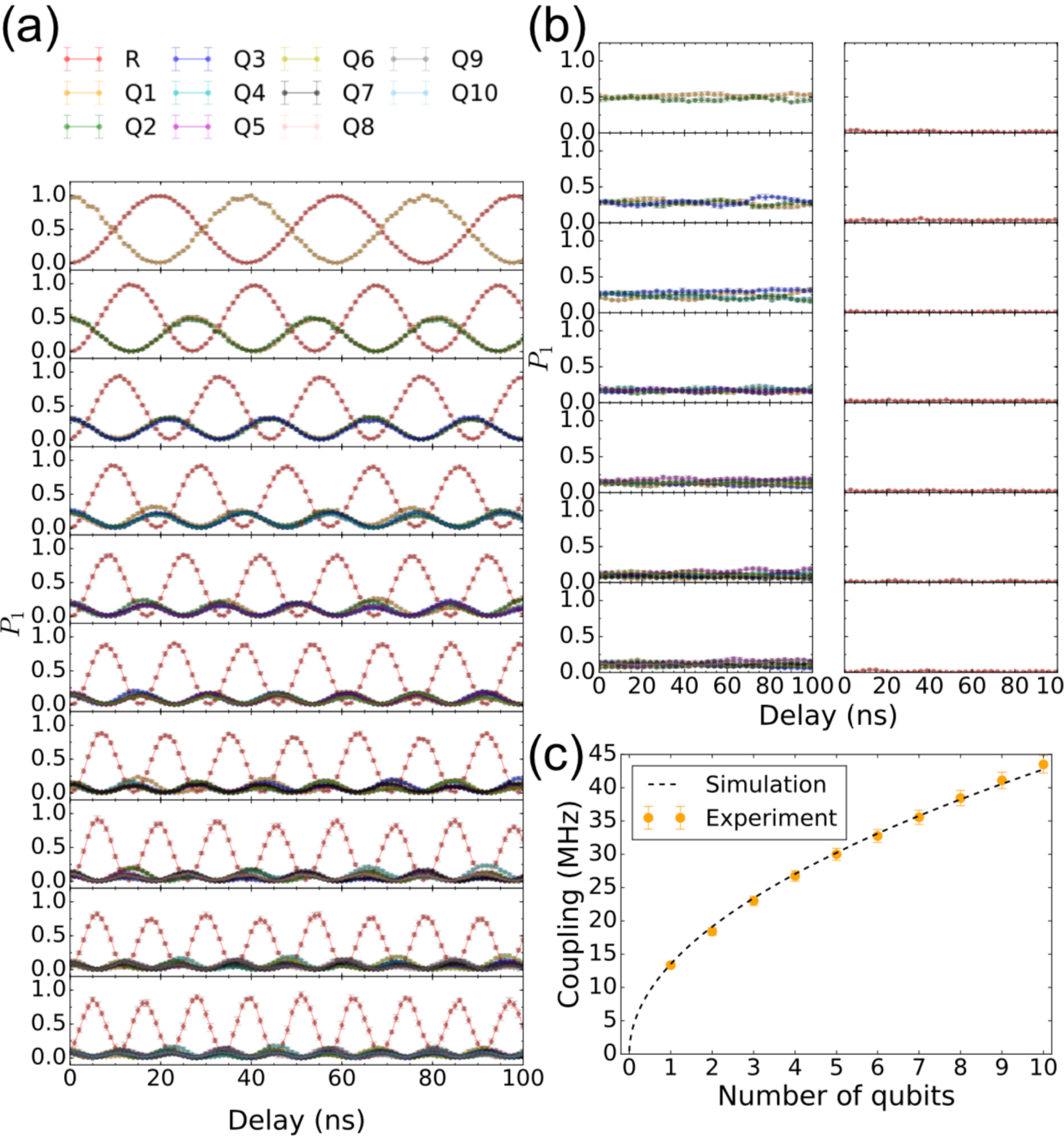}
\caption{\label{figure2} (a) Collective swapping dynamics after the $N$ qubits in superradiance are tuned into resonance with the resonator in vacuum. Shown are the population for the $|1\rangle$ state, $P_1$, of each qubit and that for the first excited state of resonator $R$, measured as functions of the swapping time using the pulse sequence shown in Fig.~1(b). For $N=10$, we infer the resonator population by subtracting the $P_1$ sum of all 10 qubits from unity, instead of directly measuring $R$ using Q$_0$ and an iSWAP gate.
(b) Collective swapping dynamics between the $N$ qubits in subradiance and $R$. $P_1$ of each qubit is measured to be around $1/N$ (left) and that for the first excited state of $R$ remains almost zero (right), where the observable small fluctuations are likely due to the slight inhomogeneity in $g_j$ and the qubit-qubit crosstalk couplings, typically less than 0.2~MHz, that are not taken into account.
(c) Collective coupling strength (dots) versus $N$ obtained by fitting the oscillation curves in (a) for the Rabi frequency. Line is a fit according to the $\sqrt{N}$ scaling.}
\end{center}
\end{figure}

We start with preparing the single-excitation superradiant state of
$N$ identical qubits $\ket{B_{N}}=\frac{1}{\sqrt{N}}\sum_{j=1}^{N}\ket{0_{1}0_{2}...1_{j}...0_{N}}$
using the pulse sequence illustrated in Fig.~1(b). After initializing
all qubits in the ground state $|0_{1}0_{2}...0_{N}\rangle$ and resonator
$R$ in vacuum, we excite Q$_{0}$ to $|1\rangle$ using a $\pi$
pulse and then tune it into resonance with $R$ using a rectangular
Z pulse for an iSWAP, which swaps the excitation quantum (i.e., a
microwave photon) into $R$. 
we apply rectangular Z pulses to all $N$ qubits for them to be on
resonance with $R$ for an appropriate duration, so that the photon
in $R$ is equally distributed among the $N$ qubits. Finally, single-qubit
phase gates, illustrated as the small rectangular Z pulses in red
in Fig.~1(b), are applied to the $N$ qubits to remove the additional
dynamical phases accumulated during the frequency tuning process,
following which we obtain $\ket{B_{N}}$ (See Supplementary Material
for quantum state tomography of $\ket{B_{N=5}}$). During the state
generation, qubits not in use are far detuned and can be ignored.

To generate the subradiant counterpart of $\ket{B_{N}}$, we apply
different single-qubit phase gates, i.e., the small rectangular Z
pulses in red in Fig.~1(b) but with different amplitudes, to locally
modify the phase of each qubit in order to generate a subradiant state
denoted as $\ket{D_{N}}=\frac{1}{\sqrt{N}}\sum_{j=1}^{N}e^{-i\phi_{j}}\ket{0_{1}0_{2}...1_{j}...0_{N}}$,
where $\phi_{j}=j\frac{2\pi}{N}$ \cite{key-31}. There are $N-2$
other subradiant states by replacing $\phi_{j}$ with $m\phi_{j}$
$(m=2,3,...,N-1)$, which can be generated with the same technique.
Here we focus on $|D_{N}\rangle$ for simplicity. The coupling strength
between $\ket{D_{N}}$ and the $N$-qubit ground state $|0_{1}0_{2}...0_{N}\rangle$
is zero since they have different exchange symmetries and the interaction
Hamiltonian retains this symmetry \cite{key-32}. However, these subradiant
states can be excited to states with the same symmetry.

\begin{figure}[t]
\begin{center}
\includegraphics[width=3.4in]{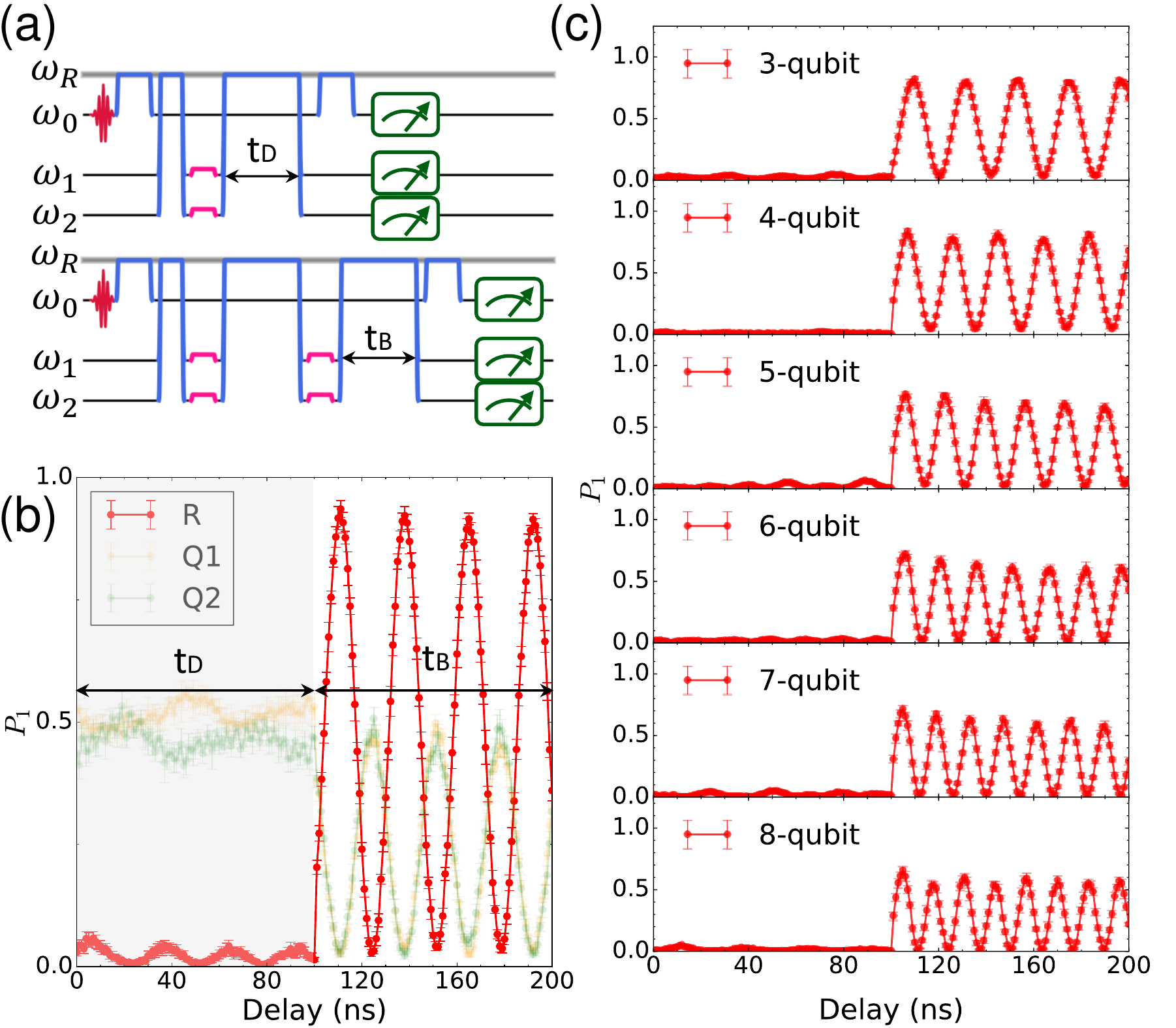}
\caption{\label{figure3} (a) Pulse sequences used to implement the switching operation from subradiance to superradiance for $N=2$. The first set of single-qubit phase gates (both the top and bottom panels), small rectangular Z pulses (red), is for generating subradiant states while the second set (the bottom panel) is inserted and configured for switching to superradiance. The collective swapping process is witnessed by tuning the $N$ qubits into resonance with $R$ using a set of big rectangular pulses (blue) followed by simultaneous readout of the $N$ qubits and $R$. (b) Experimental results using the pulse sequences in (a) as indicated. The appearance of Rabi oscillations at 100~ns signifies the entrance from subradiance to superradiance. (c)~Experimental results of the resonator photon number as function of the swap time using pulse sequences similar to those in (a), which demonstrate controllable switching operations from subradiance to superradiance with $N$ from 3 to 8.}
\end{center}
\end{figure}

\begin{figure}[htb]
\begin{center}
\includegraphics[width=3.4in]{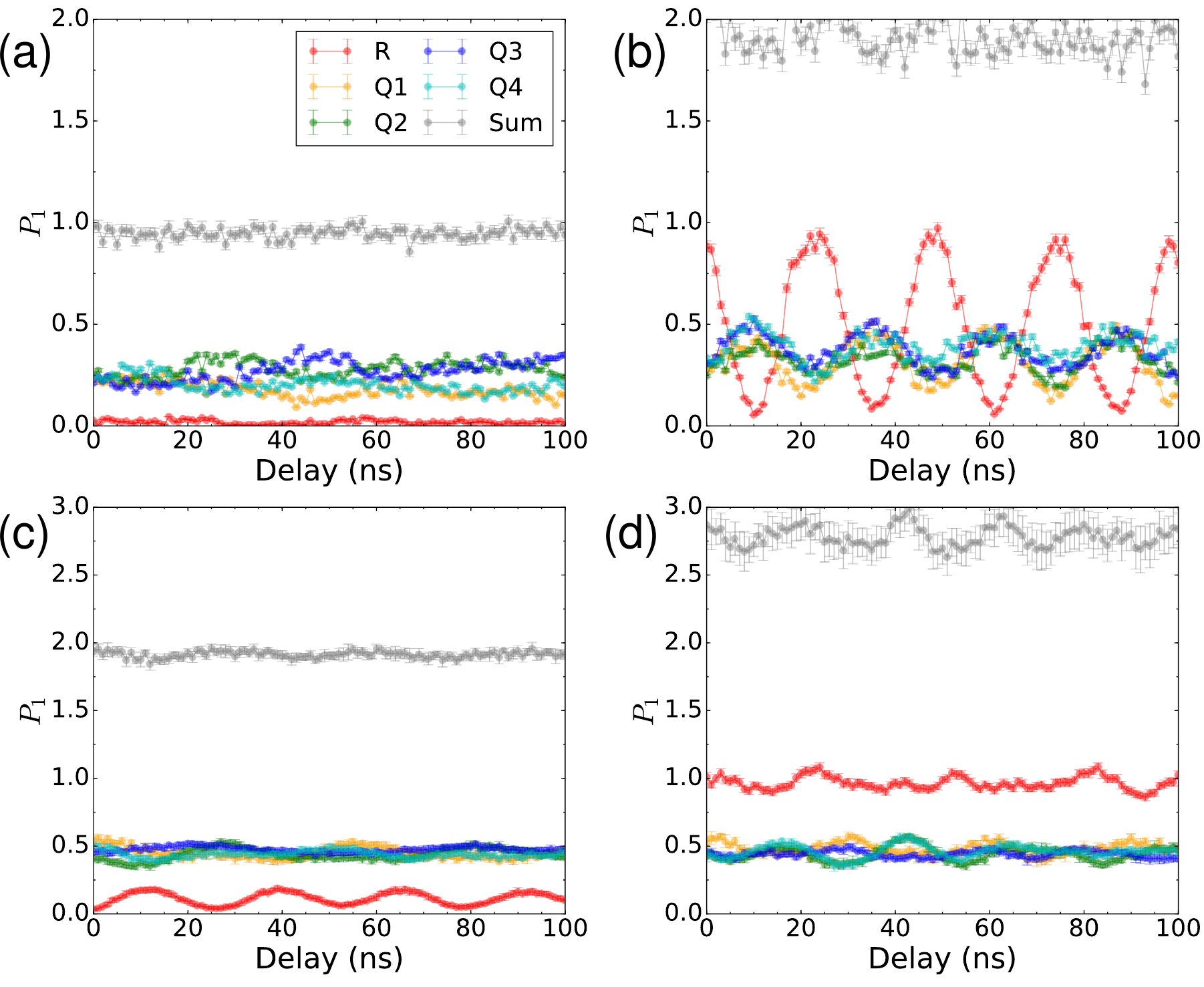}
\caption{\label{figure4} (a) Collective swapping dynamics for the single-excitation subradiant state after the 4 qubits are tuned into resonance with the resonator in vacuum. Shown are the population for the $|1\rangle$ state, $P_1$, of each of qubit and that for the first excited state of resonator $R$, measured as functions of the swapping time using the pulse sequence shown in Fig.~1(b). Variations in $P_1$ are likely due to the slight inhomogeneity in $g_j$ and the qubit-qubit crosstalk couplings that are not taken into account.
(b) Collective swapping dynamics showing the well-defined Rabi oscillations for the single-excitation subradiant state after the 4 qubits are tuned into resonance with $R$ that initially hosts a single photon.
(c) Collective swapping dynamics for the 4 qubits in $\ket{\psi_{S}}$ while $R$ is initialized in vacuum.
(d) Collective swapping dynamics for the 4 qubits in $\ket{\psi_{S}}$ while $R$ initially hosts a single photon.
}
\end{center}
\end{figure}

The most striking difference between superradiance and subradiance lies in their radiation properties, which can be witnessed by probing the the collective swapping dynamics between the $N$ qubits and $R$. Rabi oscillations in both the resonator photon number and the $|1\rangle$-state population, $P_1$, of each qubit are well observed for the superradiant states as shown in Fig.~2(a), while the photon number remains nearly zero and $P_1$ maintains almost a constant around $1/N$ for the subradiant states as shown in Fig.~2(b). Based on fittings to the Rabi oscillation curves for the superradiant states with up to 10 qubits, we extract the Rabi frequency as function of $N$, which validates the $\sqrt{N}$-scaling enhancement of the collective coupling strength in Fig.~2(c). Therefore, the radiation rate of superradiance increases with a factor $\propto\sqrt{N}$ while that of subradiance is always negligible.

As proposed by Scully \cite{key-33}, the radiation properties of
superradiance and subradiance have applications in quantum information
processing, where superradiance can be used to speed up reading/writing
and subradiance is for storing quantum information. For the proposal
to be viable, one needs to be able to switch between superradiant
and subradiant states in a controllable manner, which can be accomplished
by applying a series of single-qubit phase gates in our experiment.
As demonstrated for $N=2$ in Figs.~3(a) and 3(b), we first prepare
a $N=2$ subradiant state which stores a photon (i.e., the piece of
quantum information) for 100~ns. As such the resonator photon number
remains nearly zero and $P_{1}$ of the two qubits stay around $1/2$
during the subsequent collective swapping dynamics. At $t=100$ ns,
we pause the swapping process by detuning the two qubits from resonator
$R$, immediately apply single-qubit phase gates to switch the subradiant
to superradiant state, and then resume the collective swapping dynamics.
As expected from the radiation properties of subradiance/superradiance,
the resonator photon number 
$P_{1}$ of the two qubits oscillate again 
starting from 100 ns, yielding Rabi oscillations which signify the entrance
to superradiance. In Fig.~3(c) we perform similar switching operations
for $N$ up to 8 qubits. It is seen that, as $N$ increases, the Rabi
oscillation between the $N$ qubits and $R$, which can be used for
reading/writing, speeds up by a factor of $\sqrt{N}$ although the
efficiency, which is related to the oscillation amplitude, actually
decreases due to the inhomogeneous crosstalks between the qubits during
the subradiant stage.

One major challenge in quantum information processing is the unavoidable
decoherence due to the interaction between the quantum system and
the environment. A possible prescript to cure this problem is the
decoherence-free subspace in which the effect of the environmental
noise is minimized \cite{key-10,key-11}. The flexibility of our device
allows us to further investigate the radiation properties of the collective
states in subradiant subspaces, which are candidates for the decoherence-free
subspace. Here we show the experimental data of the collective interaction
between a group of 4 qubits in the single-excitation subradiant state
and $R$. No Rabi oscillation is observed in Fig.~4(a), which indicates
that the subradiant state is decoupled from $R$ in vacuum, i.e.,
the subradiant state cannot emit a photon to $R$. However, after
injecting an additional photon into $R$ (See Supplementary Material
for details), we observe Rabi oscillations as shown in Fig.~4(b),
i.e., this second photon can still be collectively absorbed by the
qubits in subradiance. 

A particularly interesting candidate for the noiseless quantum code
is the singlet state with even number of qubits \cite{key-10}. In
contrast to the single-excitation subradiant state, these states can
neither absorb nor emit photons, although half of the qubits are excited.
Here we generate such a state of 4 qubits by implementing two parallel
$\sqrt{\textrm{iSWAP}}$ gates~\cite{key-34} to obtain two Einstein-Podolsky-Rosen
qubit pairs in the form of $(\ket{10}-\ket{01})/\sqrt{2}$, and the
4-qubit state can be written as $\ket{\psi_{S}}=(\ket{1100}+e^{i\pi}\ket{0110}+e^{i2\pi}\ket{0011}+e^{i3\pi}\ket{1001})/2$.
We show the collective swapping dynamics of $\ket{\psi_{S}}$ for
the case of $R$ in vacuum in Fig.~4(c) and that of $R$ hosting
a single photon in Fig.~4(d). Despite the small oscillations which
are less than 10\% of the two excitation quanta hosted in the joint
system, no significant exchange of excitations between the 4 qubits
in $\ket{\psi_{S}}$ and $R$ is observed, no matter whether $R$
is in vacuum or not. The remaining exchange of photons between the
qubits and $R$ is due to the inhomogeneity of the coupling strengths
$g_{j}$, which renders $|\psi_{S}\rangle$ an imperfect subradiant
state. 

In conclusion, we have deterministically generated multiqubit superradiant/subradiant states and demonstrated controllable processing using phase gates which switch the subradiant states to their superradiant counterparts in a 10-qubit superconducting circuit. Our observation verifies the $\sqrt{N}$-scaling enhancement of the coupling strength between the superradiant states and the resonator, as well as the inhibited interaction between subradiant states and the resonator. We have also generated the singlet state of 4 qubits with 2 excitations, a possible candidate for the noiseless quantum code, which neither absorbs nor emits photons. Therefore, our experiment represents a step forward in coherent control of collective radiation and has promising applications in quantum information processing.

\noindent{\textbf{Acknowledgments}}.\noindent{
This work was supported by the National Basic Research Program
of China (Grants No.~2017YFA0304300, No.~2016YFA0300600 and No.~2018YFA0307200), the
National Natural Science Foundations of China (Grants No.~11725419, No.~11434008 and No.~11874322), and Strategic Priority Research Program of Chinese Academy of Sciences (Grant No. XDB28000000).
Devices were made at the Nanofabrication Facilities at Institute of Physics in Beijing and National
Center for Nanoscience and Technology in Beijing. }

\end{document}